\documentstyle[12pt,fleqn]{article}
\newcommand{\bea}{\begin{eqnarray}}
\newcommand{\eea}{\end{eqnarray}}

\begin{document}

\title{\Large \bf Newtonian versus relativistic nonlinear cosmology}
\author{Hyerim Noh\footnote{\rm Korea Astronomy and Space
                            Science Institute, Daejon; Koreahr@kasi.re.kr}
        \ \ and \
        Jai-chan Hwang\footnote{\rm Department of Astronomy and Atmospheric
        Sciences,
        Kyungpook National University, Taegu, Korea; jchan@knu.ac.kr}
        }
\date{}
\maketitle
\parindent = 2em

\begin{abstract}
\baselineskip 18pt

{\bf Both for the background world model and its linear
perturbations Newtonian cosmology coincides with the zero-pressure
limits of relativistic cosmology. However, such successes in
Newtonian cosmology are not purely based on Newton's gravity, but
are rather guided ones by previously known results in Einstein's
theory. The action-at-a-distance nature of Newton's gravity requires
further verification from Einstein's theory for its use in the
large-scale nonlinear regimes. We study the domain of validity of
the Newtonian cosmology by investigating weakly nonlinear regimes in
relativistic cosmology assuming a zero-pressure and irrotational
fluid. We show that, first, if we ignore the coupling with
gravitational waves the Newtonian cosmology is exactly valid even to
the second order in perturbation. Second, the pure relativistic
correction terms start appearing from the third order. Third, the
correction terms are independent of the horizon scale and are quite
small in the large-scale near the horizon. These conclusions are
based on our special (and proper) choice of variables and gauge
conditions. In a complementary situation where the system is weakly
relativistic but fully nonlinear (thus, far inside the horizon) we
can employ the post-Newtonian approximation. We also show that in
the large-scale structures the post-Newtonian effects are quite
small. As a consequence, now we can rely on the Newtonian gravity in
analyzing the evolution of nonlinear large-scale structures even
near the horizon volume.}

\end{abstract}



\baselineskip 16pt \vskip .5cm
%
%
{\it 1. Introduction:} In order to interpret results from Einstein's
gravity theory {\it properly} we often need corresponding results in
Newton's theory. On the other hand, in order to use results from
Newton's gravity theory {\it reliably} we need confirmation from
Einstein's theory. The observed large-scale structures show
nonlinear processes are working. Currently, studies of such
structures are mainly based on Newtonian physics in both analytical
and numerical approaches. One may admit its incompleteness as the
simulation scale becomes large because, first, Newton's gravity is
an action-at-a-distance, i.e., the gravitational influence
propagates instantaneously thus violating causality. Second,
Newton's theory is ignorant of the presence of horizon where the
relativistic effects are supposed to dominate. One other reason we
may add is that Einstein's gravity apparently has quite different
structure from Newton's one. The causality of gravitational
interactions and consequent presence of the horizon in cosmology are
naturally taken into account in the relativistic gravity theory.

In the literature, however, independently of such possible
shortcomings of Newton's gravity in the cosmological situation, the
physical size of Newtonian simulation, in fact, has already reached
the Hubble horizon scale. Common excuses often made by people
working in this active field of large-scale numerical simulation
are, first, in the small scale one may rely on Newton's theory and,
second, as the scale becomes large the large-scale distribution of
galaxies looks homogeneous. {\it If} the deviation from homogeneity
is small (linear) Einstein's gravity gives {\it the same} result as
the Newtonian one. Presence of large-scale homogeneity, although
difficult to verify observationally, is in fact a crucially
important assumption in currently popular cosmology. In order to
have proper confirmation, however, we still need to investigate
Einstein's case in the nonlinear or weakly nonlinear situations.
While the general relativistic cosmological simulation is currently
not available, in this work, we will shed light on the situation by
a perturbative study of the nonlinear regimes assuming zero-pressure
and irrotational fluid in Einstein's gravity. This allows us to
investigate the similarity and difference between the two gravity
theories in the weakly nonlinear regimes in cosmological situation.
We will show that even to the second order in perturbations, except
for the coupling with gravitational waves, Einstein's gravity gives
{\it the same} results known in Newton's theory and the pure
relativistic corrections appearing in the third order perturbations
are independent of the horizon and are small. We also present a
complementary approach using the post-Newtonian approximation which
can handle weakly relativistic (thus, far inside the horizon) but
fully nonlinear situation. We show that the first-order
post-Newtonian corrections are again quite small. Thus, now our
relativistic analysis {\it assures} that Newton's gravity is
practically reliable even in the weakly nonlinear regimes in
cosmology. We set $c \equiv 1$.

\vskip .5cm
%
%
{\it 2. Nonlinear equations:} We start from the completely nonlinear
and covariant equations \cite{covariant}. We need the energy
conservation equation and the Raychaudhury equation. In a
zero-pressure medium without rotation we have \cite{covariant} \bea
   & & \tilde {\dot {\tilde \mu}} + \tilde \mu \tilde \theta = 0, \quad
       \tilde {\dot {\tilde \theta}} + {1 \over 3} \tilde \theta^2
       + \tilde \sigma^{ab} \tilde \sigma_{ab}
       + 4 \pi G \tilde \mu - \Lambda = 0,
   \label{covariant-eqs}
\eea where $\Lambda$ is the cosmological constant; $\tilde \mu$ is
the energy density, $\tilde \theta \equiv \tilde u^a_{\;\; ;a}$ is
the expansion scalar with $\tilde u_a$ the fluid four-vector, and
$\tilde \sigma_{ab}$ is the shear tensor; $\tilde {\dot {\tilde
\mu}} \equiv \tilde \mu_{,a} \tilde u^a$ is the covariant
derivatives along $\tilde u^a$. Tildes indicate the covariant
quantities. By combining these equations we have \bea
   & & \left( \frac{\tilde {\dot {\tilde \mu}}}{\tilde \mu}
       \right)^{\tilde \cdot}
       - \frac{1}{3}
       \left( \frac{\tilde {\dot {\tilde \mu}}}{\tilde \mu} \right)^2
       - \tilde \sigma^{ab} \tilde \sigma_{ab}
       - 4 \pi G \tilde \mu
       + \Lambda
       = 0.
   \label{covariant-eq3}
\eea These equations are {\it fully nonlinear} and covariant. To the
second and higher order perturbations we also need the momentum
constraint part of Einstein's equation.

As the metric we take \bea
   & & ds^2
       = - a^2 \left( 1 + 2 \alpha \right) d \eta^2
       - 2 a^2 \beta_{,\alpha} d \eta d x^\alpha
       + a^2 \left[ g^{(3)}_{\alpha\beta} \left( 1 + 2 \varphi \right)
       + 2 \gamma_{,\alpha|\beta} \right] d x^\alpha d x^\beta,
   \label{metric}
\eea where $\alpha$, $\beta$, $\gamma$ and $\varphi$ are spacetime
dependent perturbed-order variables. Spatial indices of perturbed
order variables are based on $g^{(3)}_{\alpha\beta}$, and a vertical
bar indicates the covariant derivative based on
$g^{(3)}_{\alpha\beta}$. We {\it ignored} the transverse vector-type
perturbation and transverse-tracefree tensor-type perturbation
variables. In this perturbation study we will consider the {\it
scalar-type} perturbations up to third order in the {\it flat}
Friedmann background without pressure. The vector-type perturbation
has only a decaying solution in expanding medium. The presence of
gravitational waves will cause couplings with the scalar-type
perturbation to the second and higher orders in perturbations, see
\cite{second-order,third-order}. The presence of gravitational waves
can be regarded as a pure relativistic effect even to the linear
order.

\vskip .5cm
%
%
{\it 3. Background world model:} To the background order, we have
$\tilde \mu = \mu$ and $\tilde \theta = 3 {\dot a \over a}$ where
$a(t)$ is the scale factor, and an overdot indicates the time
derivative based on the background proper-time $t$. Equation
(\ref{covariant-eqs}) gives \bea
   & & \dot \mu + 3 {\dot a \over a} \mu = 0, \quad
       3 {\ddot a \over a} + 4 \pi G \mu - \Lambda = 0.
\eea This was first derived based on Einstein's gravity by Friedmann
in 1922 \cite{Friedmann-1922}, and the Newtonian study followed
later by Milne and McCrea in 1934 \cite{Milne-1934}. In the
Newtonian context $\mu$ can be identified with the mass density
$\varrho$.

\vskip .5cm
%
%
{\it 4. Linear-order perturbations:} To the linear-order
perturbations in the metric and energy-momentum variables, we
introduce \bea
   & & \tilde \mu \equiv \mu + \delta \mu, \quad
       \tilde \theta \equiv 3 {\dot a \over a} + \delta \theta.
   \label{perturbations}
\eea To the linear order we {\it identify} \bea
   & & \delta \mu \equiv \delta \varrho, \quad
       \delta \theta \equiv {1 \over a} \nabla \cdot {\bf u},
   \label{identification}
\eea where $\delta \varrho$ and ${\bf u}$ are the perturbed mass
density and the peculiar velocity in Newtonian context. In all our
relativistic (nonlinear) perturbation analyses we take the temporal
comoving gauge (which together with the irrotational condition gives
$\tilde u_\alpha = 0$ for the fluid four-vector $\tilde u_a$) and
the spatial $\gamma = 0$ gauge \cite{NL}. As these gauge conditions
fix the gauge modes completely, all the remaining variables are
equivalently gauge-invariant to {\it all} orders in perturbations;
for technical details, see Section VI.C of \cite{NL} where an
explicitly gauge-invariant combination $\delta \mu_v$ which is the
same as the $\delta \mu$ in the temporal comoving gauge and the
spatial $\gamma = 0$ gauge can be found in eq.\ (282). These gauge
conditions and our choice of the perturbation variables are
crucially important to make our conclusions.

To the linear order the perturbed part of eq.\ (\ref{covariant-eq3})
gives \bea
   & & \ddot \delta + 2 {\dot a \over a} \dot \delta - 4 \pi G \mu \delta
       = 0,
   \label{pert}
\eea which is the well known density perturbation equation in both
relativistic and Newtonian contexts; we set $\delta \equiv {\delta
\mu / \mu}$. This equation was first derived based on Einstein's
gravity by Lifshitz in 1946 \cite{Lifshitz-1946}, and the Newtonian
study followed later by Bonnor in 1957 \cite{Bonnor-1957}.

It is curious to notice that in both the expanding world model and
its linear structures the first studies were made in the context of
Einstein's gravity (Friedmann 1922; Lifshitz 1946), and the much
simpler and, in hindsight, more intuitive Newtonian studies followed
later (Milne and McCrea 1934; Bonnor 1957). Perhaps these historical
developments reflect that people did not have confidence in using
Newton's gravity in cosmology before the result was already known
in, and the method was ushered by, Einstein's gravity.  It may be
also true that only after having a Newtonian counterpart we could
understand better what the often arcane relativistic analysis shows.
It would be fair to point out, however, that the ordinarily known
Newtonian cosmology (both for the background world model and its
linear perturbations) is not purely based on Newton's gravity, but
is a guided one by Einstein's theory \cite{Layzer-1954}. In the
cosmological context Newtonian gravity is known to be incomplete and
inconsistent; these are due to lack of boundary condition at spatial
infinity and the action-at-a-distance nature of Newton's gravity.
{}For the second-order perturbations, currently we only have the
Newtonian result known in the literature. Thus, the result only
known in Newton's gravity still awaits confirmation from Einstein's
theory. Here, we are going to fill the gap by presenting the much
needed relativistic confirmation to the second order and the pure
general relativistic corrections start appearing from the third
order \cite{second-order,third-order}.

\vskip .5cm
%
%
{\it 5. Second-order perturbations:} Even to the second order we
introduce perturbations as in eq.\ (\ref{perturbations}), and {\it
take} the same identifications made in eq.\ (\ref{identification}).
To the second order the perturbed part of eq.\ (\ref{covariant-eq3})
gives \cite{NL,second-order} \bea
   & & \ddot \delta + 2 {\dot a \over a} \dot \delta - 4 \pi G \mu \delta
       = - {1 \over a^2} {\partial \over \partial t}
       \left[ a \nabla \cdot \left( \delta {\bf u} \right) \right]
       + {1 \over a^2} \nabla \cdot \left( {\bf u} \cdot
       \nabla {\bf u} \right),
   \label{ddot-delta-eq-2nd}
\eea where the second-order terms are in the right hand side. {\it
Exactly the same} equation also follows from Newton's theory
\cite{Peebles-1980}. Although we identified the relativistic density
and velocity perturbation variables we {\it cannot} identify a
relativistic variable which corresponds to the Newtonian
gravitational potential to the second order \cite{second-order}.
This may not be surprising because Poisson's equation indeed reveals
the action-at-a-distance nature and the static nature of Newton's
gravity theory compared with Einstein's gravity. In the Newtonian
context eq.\ (\ref{ddot-delta-eq-2nd}) is valid to {\it fully
nonlinear} order.

\vskip .5cm
%
%
{\it 6. Third-order perturbations:} Since the zero-pressure
Newtonian system is exact to the second order in nonlinearity, all
non-vanishing third and higher order perturbation terms in the
relativistic analysis can be regarded as the pure relativistic
corrections. We use the same {\it identification} made in eq.\
(\ref{identification}) to be  valid even to the {\it third order},
and will {\it take} the consequent additional third order terms as
the pure relativistic corrections. To the third order the perturbed
part of eq.\ (\ref{covariant-eq3}) gives \cite{third-order} \bea
   & & \ddot \delta + 2 {\dot a \over a} \dot \delta
       - 4 \pi G \mu \delta
       = - {1 \over a^2} {\partial \over \partial t}
       \left[ a \nabla \cdot \left( \delta {\bf u} \right) \right]
       + {1 \over a^2} \nabla \cdot \left( {\bf u}
       \cdot \nabla {\bf u} \right)
   \nonumber \\
   & & \qquad
       + {1 \over a^2} {\partial \over \partial t}
       \left\{ a \left[ 2 \varphi {\bf u}
       - \nabla \left( \Delta^{-1} X \right) \right] \cdot \nabla \delta
       \right\}
       - {4 \over a^2} \nabla \cdot \left[ \varphi
       \left( {\bf u} \cdot \nabla {\bf u}
       - {1 \over 3} {\bf u} \nabla \cdot {\bf u} \right) \right]
   \nonumber \\
   & & \qquad
       + {2 \over 3 a^2} \varphi
       {\bf u} \cdot \nabla \left( \nabla \cdot {\bf u} \right)
       + {\Delta \over a^2}
       \left[ {\bf u} \cdot \nabla \left( \Delta^{-1} X \right) \right]
       - {1 \over a^2} {\bf u} \cdot \nabla X
       - {2 \over 3a^2} X \nabla \cdot {\bf u},
   \label{u-eq-3rd}
\eea where the last two lines are pure third-order terms with \bea
   & & X \equiv 2 \varphi \nabla \cdot {\bf u}
       - {\bf u} \cdot \nabla \varphi
       + {3 \over 2} \Delta^{-1} \nabla \cdot
       \left[ {\bf u} \cdot \nabla \left( \nabla \varphi \right)
       + {\bf u} \Delta \varphi \right].
   \nonumber
\eea This extends eq.\ (\ref{ddot-delta-eq-2nd}) to the third order.
Notice that we need the behavior of $\varphi$ to the linear order
only; $\varphi$ is a perturbed part of three-space metric in eq.\
(\ref{metric}), related to the perturbed three-space curvature (in
our comoving gauge), and dimensionless. The third-order correction
terms in eq.\ (\ref{u-eq-3rd}) reveal that all of them are simply of
$\varphi$-order higher than the second-order terms. Thus, the pure
general relativistic effects are at least $\varphi$-order higher
than the relativistic/Newtonian ones in the second order. To the
linear order we have \cite{HN-Newtonian-1999} \bea
   & & \dot \varphi = 0,
\eea thus $\varphi = C ({\bf x})$ with {\it no} decaying mode. To
the linear order our $\varphi$ is related to the perturbed Newtonian
potential $\delta \Phi$ and the Newtonian peculiar velocity ${\bf
u}$ as \cite{second-order,third-order} \bea
   & & \varphi = - \delta \Phi
       + \dot a \Delta^{-1} \nabla \cdot {\bf u}.
\eea The temperature anisotropy of cosmic microwave background
radiation gives \cite{SW-1967,Smoot-etal-1992} \bea
   & & {\delta T \over T}
       \sim {1 \over 5} \varphi
       \sim 10^{-5}.
   \label{SW}
\eea Thus, $\varphi \sim 5 \times 10^{-5}$ in the large-scale limit
near horizon scale. Therefore, to the third order, the pure
relativistic corrections are {\it independent} of the horizon scale
and depend on the strength of linear order curvature perturbation
$\varphi$ {\it only}, and are small.

\vskip .5cm
%
%
{\it 7. Discussion:} In this work we show that Newtonian
cosmological perturbation equations remain valid in {\it all}
cosmological scales including the super-horizon scale to the second
order. We assumed a zero-pressure irrotational fluid and ignored the
coupling with gravitational waves. The pure general relativistic
correction terms start appearing from the third order. The third
order correction terms involve only $\varphi$ which is independent
of the horizon scale and is small in the large scale limit near
horizon. Therefore, one can now use the large-scale Newtonian
numerical simulation more reliably as the simulation scale
approaches and even goes beyond the horizon. All our results include
the cosmological constant thus relevant in currently favoured
cosmology.

The referee has raised a couple of interesting observations that our
conclusions do not refer to the averaging procedure
\cite{Ellis-fitting}, and the pure relativistic corrections start
appearing at third order do not depend on the physical scale and on
the averaging procedure. Indeed, our relativistic-Newtonian
correspondence and the pure relativistic correction terms do not
depend on scales nor on averaging procedure. We have reached our
conclusions by {\it comparing} the exact Newtonian equations with
the relativistic ones perturbed to the second and third orders {\it
without} taking any averaging procedure. Thus, our
relativistic-Newtonian correspondence to the second order and pure
relativistic correction terms to the third order are independent of
the averaging procedure. Notice, however, that we have achieved our
result by choosing special (and proper) variables in certain
(spatial and temporal) gauge conditions where all the variables have
corresponding unique gauge-invariant combination of variables.

The independence of the third order pure relativistic correction
terms from the scale (compared with the second-order terms) is a
sure surprise of our result. However, we would like to point out
that our pure relativistic correction terms in eq.\
(\ref{u-eq-3rd}), certainly depend on our identification of the
relativistic gauge-invariant combination of variables as the
Newtonian ones to the third order made in eq.\
(\ref{identification}); this point was emphasized above eq.\
(\ref{u-eq-3rd}). Thus, if we take other identification (of the
relativistic variables and gauges) as the Newtonian ones we could
end up with correction terms which differ from our result. Based on
our successful and clear identification with exact
relativistic-Newtonian correspondence to the second order we believe
(therefore, {\it propose}) the same identification to be valid to
the third order, and {\it suggest} the third order correction terms
in eq.\ (\ref{u-eq-3rd}) as the pure general relativistic effects
(based on our identification of the variables).

The roles of tensor-type perturbation (gravitational waves) are
studied in \cite{second-order,third-order}; vector-type perturbation
(rotation) is not important because it always decays in the
expanding phase. Why Newtonian cosmology, despite its
action-at-a-distance nature, still gives the same relativistic
results even to the second-order perturbation in {\it all} scales,
leaves room for further clarification.  Also, it would be
interesting to find cosmological situations where the pure general
relativistic correction terms in eq.\ (\ref{u-eq-3rd}) could have
observationally distinguishable consequences.

Consistency of the Newtonian (nonlinear) cosmology with the
Newtonian limit of the post-Newtonian approximation of general
relativity was also reported in \cite{Kofman-Pogosyan-1995}. In
fact, it is well known that the Newtonian hydrodynamic equations
naturally appear in the zeroth post-Newtonian order of Einstein's
gravity \cite{Chandrasekhar-1965}. In \cite{Kofman-Pogosyan-1995} it
was shown that it is essential to keep the magnetic part of Weyl
tensor in order to properly recover even the Newtonian limit in the
post-Newtonian approach. In making our proof of the
relativistic-Newtonian correspondence to the second order we assumed
irrotational and zero-pressure conditions but have not imposed any
condition on the magnetic part of the Weyl tensor; for a study based
on the covariant equations ignoring the latter quantity, see
\cite{Elst-Ellis-1998}. In fact, the magnetic part of Weyl tensor
does not vanish even to the linear order in perturbations: this
quantity valid to the second order is presented in eq.\ (96) of
\cite{NL}.

Our nonlinear perturbation approach is applicable to fully
relativistic regimes including the super-horizon scales and the
early universe. However, it is limited to the weakly nonlinear
situations where the nonlinearity is supposed to be small. A
complementary approach in handling the large-scale nonlinear
evolution in Einstein's gravity is the post-Newtonian approximation.
The post-Newtonian approach assumes $v/c$-expansion with $GM/(Rc^2)
\sim v^2/c^2 \ll 1$. Whereas our perturbation approach is applicable
in fully relativistic regime assuming weak nonlinearity, the
post-Newtonian approach is applicable in fully nonlinear regime
assuming weak (relativistic) gravity and slow motion. Thus, whereas
the perturbation approach is applicable in all scales assuming weak
nonlinearity, the post-Newtonian approach is applicable to fully
nonlinear stage but only inside the horizon. Therefore, these two
approaches are complimentary in the research of large-scale cosmic
structures. Recently, we have extended Chandrasekhar's first-order
post-Newtonian hydrodynamic approximation \cite{Chandrasekhar-1965}
to cosmological situation \cite{PN}. In \cite{PN} we show that the
first-order post-Newtonian correction terms are of order \bea
   & & \frac{GM}{Rc^2}
       \sim \frac{v^2}{c^2}
       \sim 10^{-6} - 10^{-4},
\eea compared with the Newtonian terms. Thus, although there could
appear secular effects due to time-delayed propagation of gravity,
the relativistic corrections are quite negligibly small similarly as
our third-order pure relativistic correction terms in the weakly
nonlinear regime.

Therefore, our weakly nonlinear perturbation study and the fully
nonlinear post-Newtonian study assure that in the current stage of
the large-scale structure the Newtonian hydrodynamic equations are
quite sufficient and reliable in handling the {\it dynamics}.
However, since we have {\it not} identified the relativistic
variable which corresponds to Newtonian potential to the second
order, the Newtonian equations are not supposed to be reliable where
the gravitational potential has an important role, like the
gravitational lensing.

\vskip .5cm
%
%
{\it Acknowledgments:} We wish to thank the referee for constructive
comments and suggestions. H.N. was supported by the Korea Research
Foundation Grant No. R04-2003-10004-0. J.H was supported by the
Korea Research Foundation Grant No. 2003-015-C00253.

%
%

\end{document}